\documentstyle[12pt,blois,psfig]{article}

\begin{document}
\heading{UV Properties of Primeval Galaxies}

\author{Alberto Buzzoni} {Osservatorio Astronomico di Brera}
{Via Bianchi 46 23807 Merate (Lc), Italy}

\begin{bloisabstract}
The problem of UV properties of primeval galaxies is briefly assessed
from the theoretical point of view discussing its impact on the definition
of the cosmological model.
\end{bloisabstract}

\section{Introduction}

The recent major improvements in the observation of high-redshift galaxies,
coming from HST (Williams {\it et al.} 1996; Madau {\it et al.} 1996) and 
other ground-based telescopes of the new generation (Koo {\it et al.} 1996; 
Lilly {\it et al.} 1996; Cowie and Hu 1998), have urgently called for a 
better understanding of galaxy UV spectrophotometric properties.

Optical and infrared observations of the most distant objects are in facts
inspecting their restframe ultraviolet features, so that any chance of 
successful detection of primeval galaxies should eventually rely on their
fair recognition in this photometric range.

A theoretical approach has to be preferred in this regard to match the data
since ground-based observations prevent us to explore local galaxy templates 
in the extreme UV band with comparable detail.

In this note we would briefly complement Buzzoni's (1998a,b) full analysis
assessing the general problem of the UV properties of primeval galaxies and 
its impact on the cosmological model.

\section{UV Luminosity and Star Formation Rate}

In Fig. 1 we show a synthetic c-m diagram in the $U$ vs. $U-B$ plane 
for a 4 Gyr old simple stellar population (SSP) of solar metallicity 
generating from a burst star formation according to a Salpeter IMF 
(Buzzoni 1989). This diagram might probably match the real case of a 
young elliptical galaxy as seen about $z \sim 2$.

It is evident from the figure that the SSP UV luminosity is largely dominated 
by the stars around the main sequence turn off (TO) region (note, on the 
contrary, that red giant stars are definitely too cool to give any major 
contribution at short wavelength).

This tight dependence still holds even in a more complex evolutionary scenario
dealing with a continuous star formation rate (SFR) so that a direct link 
exists between total UV luminosity of the composite stellar population and 
the relative number of young stars of higher mass (that is those with the 
hottest TO). One could therefore envisage a straightforward relationship 
between galaxy UV luminosity and {\it actual} SFR, as outlined in Fig. 2 
according to Buzzoni's (1998b) calibration.

\begin{figure}
\centerline{\psfig{bbllx=71pt,bblly=180pt,bburx=480pt,bbury=614pt,file=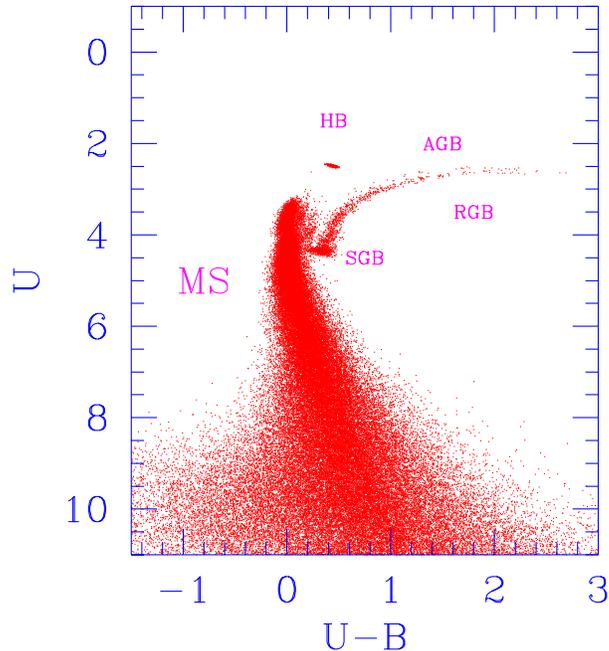,angle=0,width=8.5cm,clip=}}
\caption{Synthetic c-m diagram for a 4 Gyr Salpeter SSP of solar metallicity
(Buzzoni 1989). The random simulation consists of $10^5$ stars.
Data have been added a Poissonian photometric error such as $\sigma (U) = \pm 0.03$ 
mag at $U = +4$. The main stellar evolutionary phases are labelled in the 
figure. Note the major contribution to the total luminosity of the population
from the stars around the MS Turn Off region.}
\end{figure}

\section{The Cosmic SFR and UV Luminosity Density}

As the prevailing characteristics of the UV radiation is to track galaxy SFR, 
this leads to a quite different interpretative approach to the galaxy 
luminosity function as observed in the restframe UV range. In this case
in facts galaxy luminosity is not tracking the object size but rather its 
actual star formation activity.

A study of the luminosity function at low redshift clearly displays a change
in the Schechter fitting parameters in the sense of a steepening in the 
faint-end tail of the function as far as we move from optical to ultraviolet 
wavelengths (cf. Fig. 3).

If this trend is maintained also at high redshift, then undetected ``quiescent''
galaxies might provide a major contribution to the cosmic UV background.
Once trying to account for this large fraction of faint objects and correct
accordingly the current estimates of the cosmic UV luminosity density, 
according for instance to Madau (1997b), the final result would lead
to a measure of the cosmic SFR like in Fig. 4.

No evident signs of enhanced star formation at $z \sim 1.5$ appear from the 
figure, contrary to Madau's (1997b) original results.
The inferred SFR is a flat or decreasing function of the cosmic age, depending
on the assumed cosmological model. For an Einstein-De Sitter model, a power 
law such as ${\rm SFR} \propto t^{-1}$ seems to consistently match the observations,
while a low-density open Universe suggests a possibly constant ${\rm SFR} \sim 
0.02~[H_o/50]~M_\odot/yr$. This calls for a prevailing population of quiescent
star-forming galaxies at high-redshift, although it does eventually not
imply for those objects to be also {\it bona fide} ``primeval'' galaxies.

\begin{figure}
\centerline{\psfig{bbllx=54pt,bblly=310pt,bburx=499pt,bbury=629pt,file=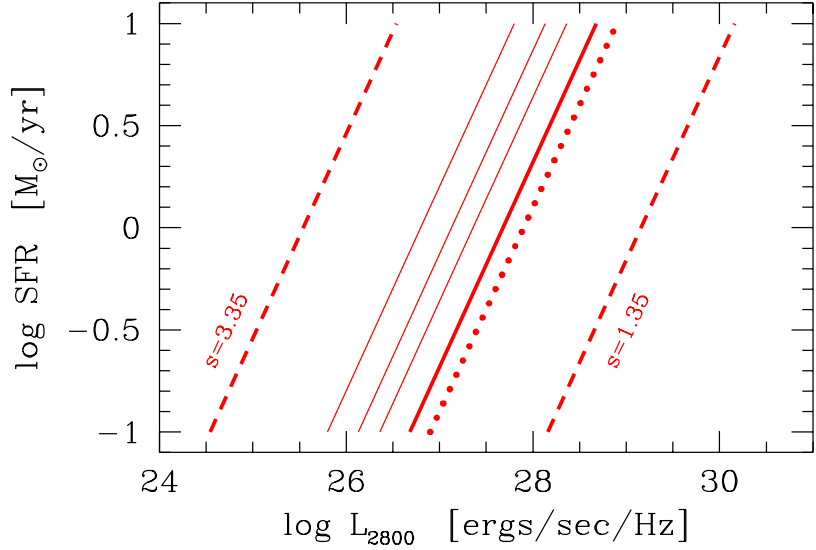,angle=0,width=11cm,clip=}}
\caption{SFR vs. UV-luminosity relationship at 2800 \AA\ according to
Buzzoni (1998b). Solid lines (from the left to the right) display the
galaxy calibration for a Salpeter IMF ($s = 2.35$) and upper star mass 
$M_{up} = 40, 60, 80,$ and $120~M_\odot$, with the last case marked in 
boldface. A change in the IMF slope (fixing $M_{up} = 120~M_\odot$ throughout)
is explored by the two dashed lines for a power-law index $s = 3.35$ and 
1.35, as labelled. Madau's (1997a) calibration for a Salpter IMF and 
$M_{up} = 125~M_\odot$ is also reported for comparison (dotted line).}
\end{figure}

\begin{figure}
\centerline{\psfig{bbllx=55pt,bblly=265pt,bburx=502pt,bbury=558pt,file=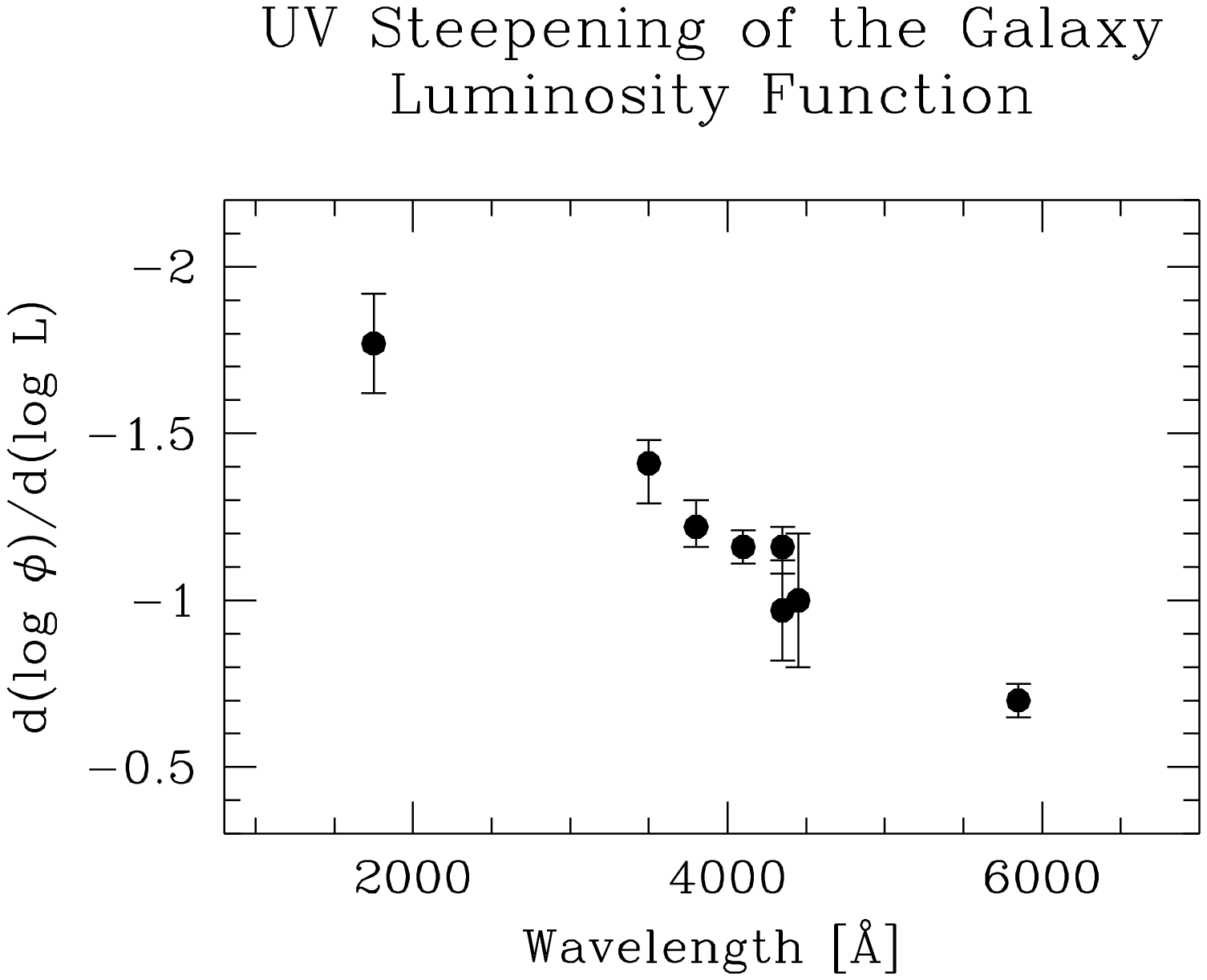,angle=0,width=11cm,clip=}}
\caption{The observed faint-end slope of the Schechter galaxy luminosity 
function ($\phi$) according to different low-redshift galaxy surveys
as collected by Buzzoni (1998b). It is evident a steepening in the 
$\partial log~\phi /\partial log~L$ derivative approaching a value of --2 in 
the UV range.}
\end{figure}

% References listed in alphabetical order ...

%
\vfill
\begin{figure}
\centerline{\psfig{bbllx=132pt,bblly=185pt,bburx=469pt,bbury=664pt,file=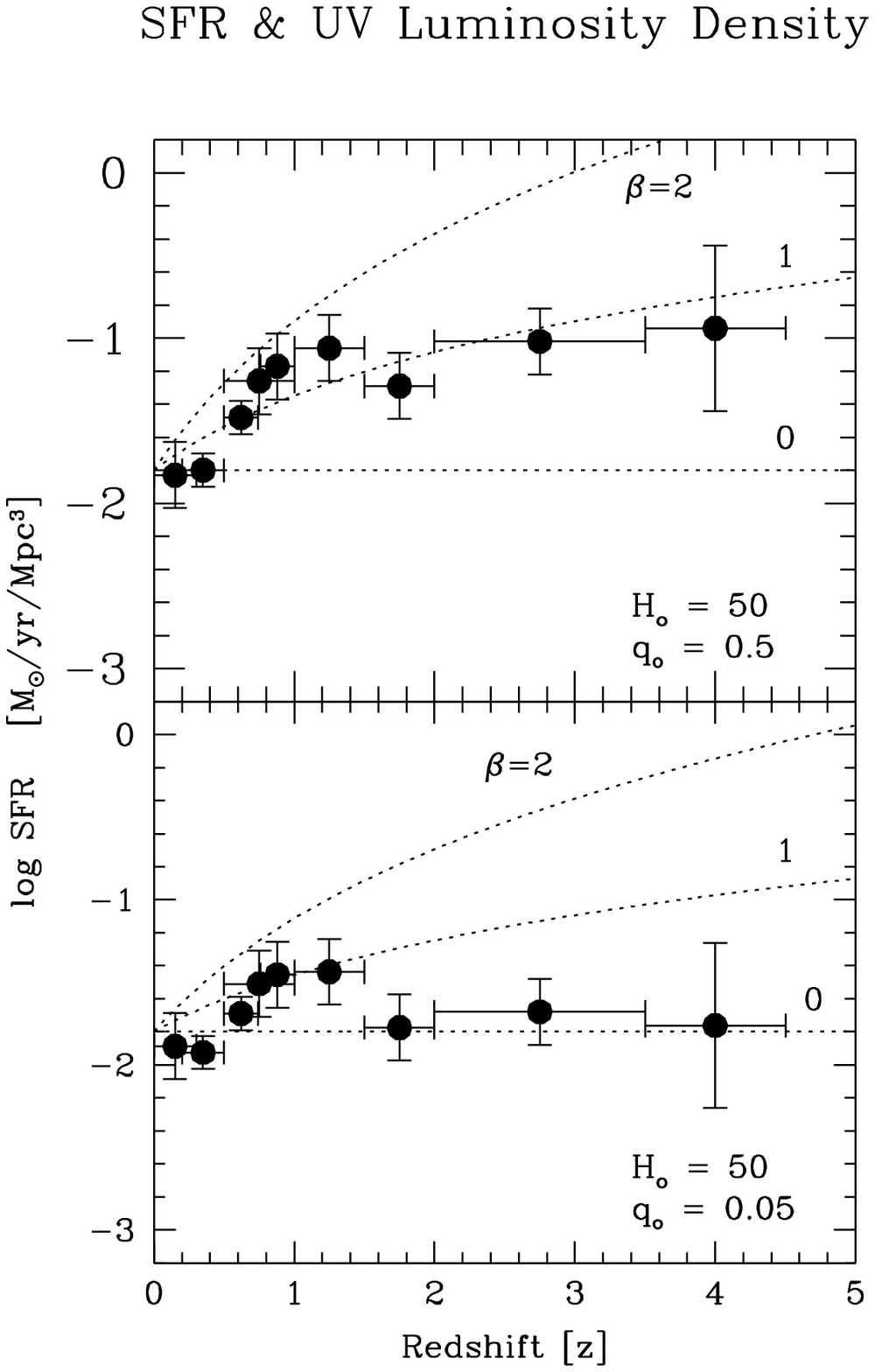,angle=0,width=8.5cm,clip=}}
\caption{The cosmic SFR as derived from the UV luminosity density after
correction for incompleteness in the faint-galaxy counts.
Two cosmological models are considered with an Einstein-De Sitter metrics
$(H_o, q_o) = (50, 0.5)$ {\it (upper panel)}, and a low-density open
model with $(H_o, q_o) = (50, 0.05)$ {\it (lower panel)}.
A power-law dependence with cosmic age such as ${\rm SFR} \propto t^{-\beta}$
with $\beta = 0, 1,$ and 2 is superposed on the plots (short-dashed lines).}
\end{figure}
\end{document}